\documentclass{article}%
\usepackage{amsmath}
\usepackage{amsfonts}
\usepackage{amssymb}
\usepackage{graphicx}%
\newcommand{\be}[1]{
\begin{eqnarray}\label{#1}}
\newcommand{\ee}{\end{eqnarray}}
\newcommand{\ci}[1]{\cite{#1}}
\newcommand{\re}[1]{(\ref{#1})}

\newcommand{\tr}{\mbox{\rm tr}}
\newcommand{\ba}{\begin{array}}
\newcommand{\ea}{\end{array}}

\newcommand{\hb}{\bar h_{v'}}
\newcommand{\Hv}{H_v}
\newcommand{\E}{E_\gamma}
\newcommand{\AS}{\text{AS}}

\begin{document}

\rightline{RUB-TPII-03/07}

\vspace*{1cm}

\begin{center}

{\Large  QCD factorization approach for rare  $\bar B^0\rightarrow  D^*\gamma$  decay  }
\\[0.5cm]

\vspace*{1cm}

\renewcommand{\thefootnote}{\fnsymbol{footnote}}
 Nikolai~Kivel\footnote{
On leave of absence from St.~Petersburg Nuclear Physics Institute,
188350, Gatchina, Russia
}
\\[3mm]
{ \it Institut f\"ur Theoretische Physik II,Ruhr-Universit\"at
Bochum, D-44780 Bochum, Germany
 }
\vspace*{1cm}

\end{center}
\renewcommand{\thefootnote}{\arabic{footnote}}

\begin{abstract}
We present the estimate of the branching ratio for the rare decay
$\bar B^0\rightarrow D^*\gamma$. We use QCD factorization approach
in order to compute the amplitude of the process. The calculation
is carried out with the leading order accuracy.
The appearing non-perturbative   matrix elements have been
estimated using the large$-N_c$ limit and QCD sum rule approach.
We obtained that $\mathcal{B}(\bar B^0\rightarrow D^*\gamma)\simeq
1.52\times 10^{-7}$.  Such value of the branching fraction is too
small in order to be measured at present experiments.
\end{abstract}
\bigskip

\newpage

\section{Introduction}

The different decay processes mediated by $b\rightarrow cd\bar u$
quark decay attract a lot of attention of both experimentalist and
theoreticians. Corresponding hadron decays include various
processes like $\bar B\rightarrow D+(\pi,\rho,K,...)$.
 There are a lot of experimental results for
the different decay modes, see for instance \cite{experiment} and
the references there. From the theoretical side the progress in
the phenomenological description of the data can be related with
the factorization approach developed last years. The factorization
theorems for the different decay channels have been discussed in
literature \cite{factorization1,factorization2,factorization3,
factorization4,factorization5,factorization6}. In the present
paper we would like to consider one particular decay mode
$B\rightarrow D^*+\gamma$ which remained beyond the considerations
mentioned above.

From the experimental point of view the process can be clearly
observed due to the higher energy of the outgoing photon
($E_\gamma\simeq 2.3$GeV). The search of this rare decay have
already been made by CLEO \cite{CLEO} and BABAR \cite{BABAR}
collaborations. Despite the process has not been observed
($\mathcal{B}(\bar B^0\rightarrow D^*\gamma)<2.5\times 10^{-5}$
\cite{BABAR}) the increasing statistics of the $B-$factories may
provide new opportunities for the better analysis. The various
existing theoretical models \cite{theor1,theor2,theor3} estimate
the branching to be of order of $10^{-6}$. Potentially, such cross
section  can be observed despite to the small value and therefore
the more qualitative theoretical analysis is desirable.

In present paper we use the factorization technique developed last
years for the heavy quarks decays in order to derive the leading
order  factorization formula for the amplitude of the process and
 estimate the branching ratio. Our presentation
is organized  as follows. Sec.~2 contains the necessary
definitions and derivation of the leading order factorization
calculations. In Sec.3 we consider the arising soft matrix
elements and construct the models for these non-perturbative
functions using large$-N_c$ limit and QCD sum rules.
 This section contains also our main results,
 the summary and discussions.

\section{The leading order amplitude}
The decay amplitude  $\bar B^0(P_B)\rightarrow D^*(P_D)\gamma(q) $
is given by the matrix element \be{Ampl} A_{D^{\ast}\gamma}  &
=&~\sqrt{4\pi\alpha}~i\int d x
e^{i(qx)}~\varepsilon_{\gamma}^{*\mu}\left\langle
P_D,\varepsilon^*_D\left\vert T\left\{
J_{\mu}^{em}(x),H_{eff}(0)\right\}  \right\vert P_B\right\rangle
 \\
& =&\frac{1}{2(q\cdot P_{D})}i\varepsilon^{\mu\nu\sigma\rho}
(\varepsilon^*_{\gamma})_\mu (\varepsilon^*_{D})_\nu
q_\sigma(P_{D})_\rho\, F_{1}+\left\{  \left(
\varepsilon^*_{\gamma}\cdot\varepsilon^*_{D}\right)
 -\frac{1}{(q\cdot P_{D})}(q\cdot\varepsilon^*_{D})(P_{D}
\cdot\varepsilon^*_{\gamma})\right\}\,  F_{2}%
\label{fmfr} \ee which is described by the two form factors
$F_{1,2}$. Here we accept standard notation
$\alpha=\frac{e^2}{4\pi}\simeq 1/137$ and $\varepsilon^*_{\gamma,
D}$ denotes photon and $D-$meson polarization vector
respectively\footnote{The antisymmetric tensor is defined as
$\varepsilon^{0123}=+1$}. The kinematics of the decay is very
simple. As usual, we choose the frame where $B-$meson is at rest.
Then the the components of the momenta read
\be{kincs} P_{B}  &
=&P_{D}+q,\,\, P^2_{B}=M^2_{B},\,\, P^2_{D}=M^2_{D},\,\, q^2=0,
\\
P_{B}  & =&M_{B}v,\,\,P_{D} =M_{D}v^{\prime},\,\,q=2E_{\gamma}\frac{n}{2},
\\
v&=&(1,0,0,0)=\frac{\bar{n}}{2}+\frac{n}{2},\,\,
 n^2=\bar n^2=0,\,\, n\cdot \bar n=2,
\\
v^{\prime}&=&\left(  \frac{M_{B}^{2}+M_{D}^{2}%
}{2M_{B}M_{D}},0,0,\frac{M_{B}^{2}-M_{D}^{2}}{2M_{B}M_{D}}\right)
=\frac{1}{x}\frac{\bar{n}}{2}+x\frac{n}{2},\, \, x=M_D/M_B,
\\
 E_{\gamma}&=&\frac{M_{B}^{2}-M_{D}^{2}}{2M_B},
\ee
where we introduced the light-cone vectors $n,\bar n$ and for arbitrary vector $a$
one has
\be{lcbasis}
a =a_+ \frac{\bar n}2+a_- \frac{ n}2+a_\bot
\ee
 Substituting the numerical values of the
heavy meson masses $M_D=2$GeV and $M_B=5.28$GeV one finds $E_{\gamma}\approx 2.3$GeV, i.e.
the photon energy is quite large.
The width is given by%
\be{width}
\Gamma_{D^{\ast}\gamma}=\frac{1}{32\pi}\frac{M_{B}^{2}-M_{D}^{2}}{M_{B}^{3}%
}\left(  ~\left\vert F_{1}\right\vert ^{2}+4\left\vert F_{2}\right\vert
^{2}\right)
\ee
Using the experimental constrain for the branching \ci{BABAR}%
\be{BABAR}
\mathcal{B}(\bar{B}^{0}\rightarrow D^{\ast0}\gamma)<2.5\times10^{-5}%
\ee%
and the lifetime $ \tau_{B^{0}}=1.536\times 10^{-12}~s $
 one can find for the combination of the form factors in \re{width}
\be{ffexp}
 \left\vert F_{1}\right\vert ^{2}+4\left\vert F_{2}\right\vert
^{2}
<1.1\times10^{-3} \left(  G_{F}\sqrt{2\pi\alpha}\right)^2,%
\ee
where the coefficient~$G_{F}\sqrt{2\pi\alpha}$ is introduced for convenience.

Our task is to compute the form factors $F_{1,2}$ in the limit
$m_b, m_c\rightarrow\infty$ with $m_c/m_b$  fixed.  The effective
Hamiltonian in the matrix element \re{Ampl} reads \be{Heff}
H_{eff}& =&\frac{G_{F}}{\sqrt{2}}V_{cb}V_{ud}^{\ast}\left[
 C_{1}~(\bar{c}b)_{V-A}(\bar{d}u)_{V-A}+
 C_{2}~\left(  \bar{d}b\right)  _{V-A}\left(  \bar
{c}u\right)  _{V-A}\right] \ee where as usually
$V-A=\gamma_\mu(1-\gamma_5)$ and the color indices are not shown
explicitly.
Let us  introduce the following parametrization for the amplitude \re{Ampl}:%
\be{amplnew}
A_{D^{\ast}\gamma}=\sqrt{M_{D}M_{B}}\sqrt{2\pi\alpha}\,\, G_{F}%
~V_{cb}V_{ud}^{\ast}~\left[
~\alpha^{\text{f}}+~\alpha^{\text{nf}}\right]  , \ee where the
coefficients $\alpha^{\text{nf}}$ and $\alpha^{\text{f}}$ are
related to the matrix elements of the  two operators in the
\re{Heff}: \be{alff} \alpha^{\text{f}}  &\displaystyle =i \int
dx~e^{i(qx)}~\varepsilon_{\gamma}^{*\mu} \left\langle
v^{\prime}\left\vert T\left\{
 J^{em}_{\mu}(x),
 C_{2} \left(\bar{d}b\right)  _{V-A}\left(  \bar{c}u \right)  _{V-A}
\right\}
\right\vert v\right\rangle,
\\
\alpha^{\text{nf}}  &\displaystyle =i \int dx~e^{i(qx)}~\varepsilon_{\gamma}^{*\mu}
\left\langle v^{\prime}\left\vert
T\left\{
 J^{em}_{\mu}(x),
 C_{1} (\bar{c}b)_{V-A}(\bar{d}u)_{V-A}
\right\} \right\vert v\right\rangle, \label{alfnf} \ee The meaning
of the superscripts "f, nf" will be explained below. In these
formulas we assume that the meson states $\vert v\rangle $, $
\langle v^{\prime}\vert$ have mass independent HQET normalization.

\begin{figure}
[ptb]
\begin{center}
\includegraphics[
height=2cm
]%
{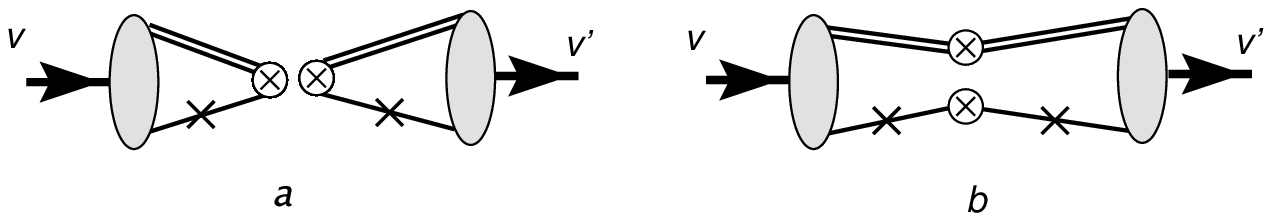}%
\end{center}
\caption{The leading order diagrams $a,\, b$ denote the graphs for the form
factors $\alpha^{\text{f}}$} and $\alpha^{\text{nf}}$ respectively.
The crossed lines denotes the emission of the photon.
\label{LOgraph}
\end{figure}

In the large mass limit the energy of the photon is also large
$E_\gamma\sim m_Q\rightarrow\infty$. The emission of such higher
energy photon is related with short distance subprocess. In some
sense, the similar situation is encountered in the case of
semi-leptonic decay $B\rightarrow\gamma l\nu$. The difference with
respect to our case is  in the more complicate structure of the
matrix element \re{Ampl}. Consider the simplest diagrams which can
contribute at the leading order Fig.\ref{LOgraph}.
The analytical expression for the case  $\alpha^\text{f}$ reads
\be{calc1} \alpha^{\text{f}}  = 4iC_{2}\int dx\,
e^{i(qx)}\left\langle v^{\prime}\left\vert e_{d}~\bar
{c}~\gamma^{\mu}P_L~u~\bar{d}(x)~\hat{\varepsilon}^*_{\gamma}\hat{\Delta}(x,0)\gamma_{\mu}P_L
~b(0)+e_{u}~\bar{d}(0)\gamma^{\mu}P_L
b~\bar{c}(0)\gamma_{\mu}P_L~\hat{\Delta}(0,x)\hat
{\varepsilon}^*_{\gamma}u(x)\right\vert v\right\rangle, \ee where
$\Delta(x,y)$ is the fermion propagator in position space,
$e_{u,d}$ are  quark charges,  and the hat denotes the
contractions with the Dirac matrices $a_\mu \gamma^{\mu}=\hat a$
and $P_{L}=\frac{1}{2}(1-\gamma_5)$. In the heavy quark limit one
performs transition from the QCD heavy quark fields to the HQET
fields:
\be{HQETf}
b(0)\simeq H_{v},~\ \bar{c}(0)\simeq \bar
{h}_{v^{\prime}} \ee Hence \be{calc2}
\alpha^{\text{f}}  \simeq 4C_{2}i\int dx~e^{i(qx)}\left\langle v^{\prime}\left\vert e_{d}%
~\hb~\gamma^{\mu}P_L~u~\bar{d}(x)~\hat{\varepsilon}^*_{\gamma}\hat{\Delta}%
(x,0)\gamma^{\mu}P_L~\Hv+
e_{u}~\bar{d}~\gamma^{\mu}P_L \Hv~\hb~\Gamma~\hat{\Delta}%
(0,x)\hat{\varepsilon}^*_{\gamma}u(x)\right\vert v\right\rangle \\
 =4iC_{2}\int dk\ e_{d}\left[  ~\hat{\varepsilon}^*_{\gamma}\hat{\Delta
}(-k+q)\gamma^{\mu}P_L\right]  _{\alpha\beta}~\int dx~e^{i(kx)}
\left\langle v^{\prime}\left\vert
\hb\gamma_{\mu}P_L u~\left[  \bar{d}_{\alpha}(x)~(\Hv)_{\beta}
\right]  \right\vert v\right\rangle +
\nonumber \\
 4iC_{2}\int dl\ e_{u}\left[  \gamma^{\mu}P_L~\hat{\Delta}(-l-q)
\hat{\varepsilon}^*_{\gamma}\right]  _{\alpha\beta}~
\int dx~e^{-i(lx)}
\left\langle v^{\prime
}\left\vert \left[  (\hb)_{\alpha}~u_{\beta}(x)\right]\bar{d}\gamma_{\mu}P_L \Hv
\right\vert v\right\rangle\, .
\label{momsp}
\ee
In the second line we performed the transition to the momentum space, indices $\alpha, \beta$
denote the spinor indices, $dk\equiv d^4k/(2\pi)^4$. To proceed further we assume that given
expression is dominated by the region, where the momenta $k$ and $l$ are soft:
\be{softmom}
k_i\sim l_i\sim \bar\Lambda,\, \, k^2\sim l^2\sim \bar{\Lambda}^2,
\ee
where $ \bar\Lambda\simeq M_Q-m_Q$ is the soft scale.
Then in that region the expressions for the matrix elements
$\langle v^{\prime}\vert...\vert v\rangle$ in \re{momsp}  are defined
only in terms of the long wave
fields and can be understood as  soft matrix elements. The expressions
in the $[...]_{\alpha\beta}$ can be simplified for the large energy
($E_\gamma\sim m_Q$):
\begin{align}
\hat{\varepsilon}^*_{\gamma}\hat{\Delta}(-k+q)\gamma^{\mu}P_L & =
\hat{\varepsilon}^*_{\gamma}\frac{i(-\hat k+\hat q)}{(-k+q)^{2}
+i\varepsilon}\gamma^{\mu}P_L
\simeq\hat{\varepsilon}^*_{\gamma}
\hat{q}\gamma^{\mu}P_L\frac{i}{\left[  -2(kq)+i\varepsilon\right]  },
\\
\gamma^{\mu}P_L ~\hat{\Delta}(-l-q)\hat{\varepsilon}^*_{\gamma}  & =\gamma^{\mu}P_L
~\frac{i(-\hat l-\hat q)}{(-l-q)^{2}+i\varepsilon}\hat{\varepsilon}^*_{\gamma}\simeq
\gamma^{\mu}P_L~\hat{q}\hat{\varepsilon}^*_{\gamma}\frac{-i}{\left[  2(lq)+i\varepsilon
\right]  }%
\end{align}%
Substituting these expressions into \re{momsp}
\begin{align}
\alpha^{\text{f}}  & =4iC_{2}\int dk\
\frac{i e_{d}
\left[  \hat{\varepsilon}^*_{\gamma}\hat{q}\gamma^{\mu}P_L \right]  _{\alpha\beta}}
{\left[  -2(kq)+i\varepsilon\right]}
~\int dx~e^{i(kx)}\left\langle v^{\prime}\left\vert \bar{h}\gamma_{\mu}P_L u~
\left[
\bar{d}_{\alpha}(x)~H_{\beta}
\right]  \right\vert v\right\rangle
\nonumber \\&
+4iC_{2}\int dl\
\frac{-i e_{u}\left[
\gamma^{\mu}P_L~\hat{q}~\hat{\varepsilon}^*_{\gamma}\right]  _{\alpha\beta}}
{\left[  2(lq)+i\varepsilon\right]  }~
\int dx~e^{-i(lx)}\left\langle v^{\prime}\left\vert \left[  \bar{h}_{\alpha
}~u_{\beta}(x)\right]  ~\bar{d}\gamma_{\mu}P_L H~\right\vert v\right\rangle
\end{align}%
Performing integrations over $dk_{-}$ and $dk_\bot$ and then over the conjugate
variables $x_+$ and $x_\bot$ (and similar for the second term with momentum $l$)
we obtain
\be{calc3}
\alpha^{\text{f}} &
\displaystyle
 =C_{2}\frac{1}{2}\int dk_{+}
\frac{
-e_{d}\tr\{
\hat{\varepsilon}^*_{\gamma}\hat{q}\gamma^{\mu}\gamma^{\rho}P_{R}%
\}}
{\left[  -2\E~k_{+}+i\varepsilon\right]  }
\int\frac{d\lambda_{1}}{2\pi}~e^{ik_+\lambda_{1}}
\left\langle v^{\prime}\left\vert \bar
{h}\gamma_\mu P_L u  ~\bar{d}\left(  \lambda_{1}~n\right)  \gamma_{\rho}%
P_{L}~H \right\vert v\right\rangle
\nonumber \\& \displaystyle
+C_{2}\frac{1}{2}\int dl_{+}\ \frac{e_{u}
\tr\{ \gamma^{\mu}%
\hat{q}~\hat{\varepsilon}^*_{\gamma}\gamma^{\rho}P_{R}\} }
{\left[
2\E~l_{+}+i\varepsilon\right]  }
\int\frac{d\lambda_{2}}{2\pi}~e^{-il_+\lambda_{2}}
\left\langle v^{\prime}\left\vert  ~\bar{h}~\gamma_{\rho}%
P_{L}u\left(  \lambda_{2}~n\right)    ~\bar{d}\gamma_\mu P_L H\right\vert
v\right\rangle.
\label{calc4}
\ee
The  formula \re{calc4} represents the form factor $\alpha^{\text{f}}$
 as a convolution of the
soft light-cone matrix elements with the expression which, obviously, is associated with
the hard coefficient function. The arguments of the fields which are not written
explicitly in the eq.\re{calc4} are set  to zero.
From the structures of the traces one observes that only the combinations
antisymmetrical with respect to exchange $\mu\leftrightarrow\rho$  survive
in the soft matrix elements. Therefore  we define
\be{softme1}
S_{d}^{[\sigma\rho]}(k_+)  =\AS \int\frac{d\lambda_{1}}{2\pi}~e^{ik_+\lambda_{1}}
\left\langle v^{\prime}\left\vert
\hb\gamma_\sigma P_L u ~\bar{d}\left(  \lambda_{1}~n\right)  \gamma_{\rho}%
P_{L}~\Hv  \right\vert v\right\rangle,
\ee
\be{softme2}
S_{u}^{[\sigma\rho]}(l_+) = \AS  \int\frac{d\lambda_{2}}{2\pi}~e^{-il_+\lambda_{2}}
\left\langle v^{\prime}\left\vert  ~\hb~\gamma_{\rho}%
P_{L}u\left(  \lambda_{2}~n\right)~\bar{d}\gamma_\sigma P_L \Hv\right\vert
v\right\rangle,
\ee
where symbol "AS" denotes antisymmetrisation with respect to indices
$\left\{\sigma,\rho\right\}$, for instance%
\be{def:as}
\text{AS}~\bar{n}^{\sigma}\varepsilon_{D}^{*\rho}=\frac{1}{2}
\left(
\bar{n}^{\sigma}\varepsilon_{D}^{*\rho}-\bar{n}^{\rho}\varepsilon_{D}^{*\sigma}%
\right)
\ee
The parametrisation of these functions can be written as\footnote{
we use notation
$
i\varepsilon_{\bot}^{\rho\sigma}=\frac12 i\varepsilon^{\rho\sigma\mu\nu}n_\mu \bar n_\nu\, .
$
}
\be{def:Uf}
S_{u}^{[\sigma\rho]}(l_+)
& =\frac{i}{2}U^{\text{f}}(l_+)~\text{AS}~\bar{n}_{\sigma}
\left\{
(\varepsilon^*_{D})_{\rho}-i\varepsilon_{{\bot}\rho\mu}\varepsilon_{D}^{*\mu}
\right\}
+\frac{i}{2}\tilde{U}^{\text{f}}(l_+)~\text{AS}~\bar{n}_{\sigma}
\left\{
(\varepsilon^*_{D})_{\rho}+i\varepsilon_{{\bot}\rho\mu}\varepsilon_{D}^{*\mu}
\right\}  ,
\ee
\be{def:Df}
S_{d}^{[\sigma\rho]}(k_+)
& =\frac{i}{2}D^{\text{f}}(k_+)~\text{AS}~\bar{n}_{\sigma}
\left\{
(\varepsilon^*_{D})_{\rho}-i\varepsilon_{{\bot}\rho\mu}\varepsilon_{D}^{*\mu}
\right\}
+\frac{i}{2}\tilde{D}^{\text{f}}(k_+)~\text{AS}~\bar{n}_{\sigma}
\left\{
(\varepsilon^*_{D})_{\rho}+i\varepsilon_{{\bot}\rho\mu}\varepsilon_{D}^{*\mu}
\right\}.
\ee

Then the final result for the form factor reads
\be{af}
\alpha^{\text{f}}=\left[
 ~(\varepsilon^*_{\gamma}\cdot\varepsilon^*
_{D})+i\varepsilon_{\perp}^{\mu\nu}(\varepsilon^*_{D})_\mu(\varepsilon^*_{\gamma})_\nu
\right]
~ie_{d}~C_{2}\, \mathcal{D}^{\text{f}}+
\left[  ~(\varepsilon^*_{\gamma}%
\cdot\varepsilon^*_{D})-
i\varepsilon_{\perp}^{\mu\nu}(\varepsilon^*_{D})_\mu(\varepsilon^*_{\gamma})_\nu
\right]  ie_{u}C_{2}~\mathcal{U}^{\text{f}},%
\ee
where we introduced the convolution integrals
\begin{equation}
\mathcal{D}^{\text{f}}=\int_{0}^{\infty}dk_{+}~\frac{D^{\text{f}}\left(
k_{+}\right)  ~}{~k_{+}},
~\ \ \mathcal{U}^{\text{f}}=\int_{0}^{\infty}%
dl_{+}\frac{\tilde{U}^{\text{f}}\left(  l_{+}\right)  }{~l_{+}}%
\label{conv:f}
\end{equation}

The similar calculation for the second form factor $\alpha^{\text{nf}}$
provides
\be{anf}
\alpha^{\text{nf}}=\left[
 ~(\varepsilon^*_{\gamma}\cdot\varepsilon^*
_{D})+i\varepsilon_{\perp}^{\mu\nu}(\varepsilon^*_{D})_\mu(\varepsilon^*_{\gamma})_\nu
\right]
~ie_{d}~C_{1}\, \mathcal{D}^{\text{nf}}+
\left[  ~(\varepsilon^*_{\gamma}%
\cdot\varepsilon^*_{D})-
i\varepsilon_{\perp}^{\mu\nu}(\varepsilon^*_{D})_\mu(\varepsilon^*_{\gamma})_\nu
\right]
ie_{u}C_{1}~\mathcal{U}^{\text{nf}}, %
\ee
where the convolution integrals
\be{conv:nf}
\mathcal{D}^{\text{nf}}=\int_{0}^{\infty}dk_{+}~\frac{D^{\text{nf}}\left(
k_{+}\right)  ~}{~k_{+}},
~\ \ \mathcal{U}^{\text{nf}}=\int_{0}^{\infty}%
dl_{+}\frac{\tilde{U}^{\text{nf}}\left(  l_{+}\right)  }{~l_{+}}%
\ee
include the contributions from the different soft matrix elements
\be{def:me1}
&\displaystyle \AS  \int\frac{d\lambda_{2}}{2\pi}~e^{-il_+\lambda_{2}}
\left\langle v^{\prime}\left\vert  ~\hb~\gamma_{\rho}%
P_{L}\Hv ~\bar{d}\gamma_\sigma P_L u\left(  \lambda_{2}~n\right) \right\vert
v\right\rangle
\nonumber \\
&
\phantom{probel} =\frac{i}{2}U^{\text{nf}}(l_+)~\text{AS}~\bar{n}_{\sigma}
\left\{
(\varepsilon^*_{D})_{\rho}-i\varepsilon_{{\bot}\rho\mu}\varepsilon_{D}^{*\mu}
\right\}
+\frac{i}{2}\tilde{U}^{\text{nf}}(l_+)~\text{AS}~\bar{n}_{\sigma}
\left\{
(\varepsilon^*_{D})_{\rho}+i\varepsilon_{{\bot}\rho\mu}\varepsilon_{D}^{*\mu}
\right\}
\label{menf1}  ,
\ee
\be{def:me2}
&\displaystyle
\AS \int\frac{d\lambda_{1}}{2\pi}~e^{ik_+\lambda_{1}}
\left\langle v^{\prime}\left\vert \hb
\gamma_\sigma P_L \Hv ~\bar{d}\left(  \lambda_{1}~n\right)  \gamma_{\rho}%
P_{L}~u  \right\vert v\right\rangle
\nonumber \\
&\phantom{probel} =\frac{i}{2}D^{\text{nf}}(k_+)~\text{AS}~\bar{n}_{\sigma}
\left\{
(\varepsilon^*_{D})_{\rho}-i\varepsilon_{{\bot}\rho\mu}\varepsilon_{D}^{*\mu}
\right\}
+\frac{i}{2}\tilde{D}^{\text{nf}}(k_+)~\text{AS}~\bar{n}_{\sigma}
\left\{
(\varepsilon^*_{D})_{\rho}+i\varepsilon_{{\bot}\rho\mu}\varepsilon_{D}^{*\mu}
\right\}.
\label{menf2}
\ee
Let us briefly comment the obtained results. We have performed the
calculation only of the leading order diagrams. The matrix elements of the
non-local four-fermion operators (\ref{softme1},\ref{softme2}) and
(\ref{menf1},\ref{menf2}) consist of the product of
two field substructures: local one and non-local one.
The non-local part is presented by the two quark fields separated by light-cone distance.
It is clear that such block is not gauge invariant and therefore the answer
is not complete. To restore the gauge invariance one has to
consider the diagrams with the emissions of the soft gluons from the active quark.
This will restore the gauge link
\be{glink}
E\left[  \lambda_{1},\lambda_{2}\right]  =P\exp\left(  ig\int_{0}%
^{1}du~(\lambda_{1}-\lambda_{2})\, A_+\left[  (u\lambda_{1}+\bar{u}\lambda
_{2})~n\right]  \right)
\ee
which connects the fields and therefore completes the
definitions of the soft operators. We do not present these details because
they are standard. One can avoid that using the light-cone gauge $A_+=0$.
Then the gauge link \re{glink} equals to one and the
 definitions (\ref{softme1},\ref{softme2}) and (\ref{menf1},\ref{menf2}) in this case
are exact.

The important question which has to be considered is
the exitance of the convolution integrals
(\ref{conv:f}) and (\ref{conv:nf}).
In order to answer it one has to consider
the next-to-leading order calculation of the amplitude or at
least the evolution kernels of the soft operators.
Moreover, such calculation is important in order to
perform the summation of large logarithms which usually  appear in the
radiative corrections.  From our calculation we observe that typical virtuality
of the hard quark is of order $\sim \bar \Lambda E_\gamma$, i.e. we computed the leading order
contribution to the so-called jet function. The loop corrections contain also corrections
from the different hard subprocess with the virtualities  of order $\sim m^2_Q$. The presence
of the two large scales unavoidably leads to large logarithms mentioned above.
In order to formulate the factorization in the general case
(i.e. valid to all orders in the QCD perturbation theory)
 it is convenient to involve the technical approach
known as soft collinear effective theory (SCET)\cite{SCET1,SCET2}.
In the present paper  we do not provide such detailed analysis and
restrict our consideration to the phenomenological estimate of the
decay width \re{width} using the leading order formulas \re{af}
and \re{anf}.  Below we consider various models for the soft
matrix elements which we need for the numerical analysis. We shall
see that these models are in agreement with the factorization,
i.e. they have the appropriate end-point behavior which makes the
convolutions integrals well defined. Of course, this is not a
proof but it can be considered as an indication that the
factorization in the case under consideration is not destroyed by the end-point
singularities.

\section{The  soft matrix elements and decay width}
Our task is to estimate the non-perturbative matrix elements
defined in the previous section. The corresponding functions
$F^{\text{f,nf}}=\{U^{\text{f,nf}},D^{\text{f,nf}},
\tilde{U}^{\text{f,nf}},\tilde{D}^{\text{f,nf}} \}$ depend on
momentum fraction of the light quark $k_{+}$, velocities $v$ and
$v'$,  and factorization scale $\mu_F$. In general one can write
 \be{me:arg}
F^{\text{f,nf}}=v_+ F^{\text{f,nf}}(k_+/v_+,v_+'/v_+, (v\cdot v'),\mu_F )
\ee
The values of the $(v\cdot v'),v_+,v'_+$ are fixed by
kinematics and we shall not consider this arguments  as an arbitrary variables.
 The factorization scale $\mu_F$ we shall assume to be of order
$\bar \Lambda E_\gamma\sim 1.5$GeV. Usually, in that case one has to consider
the resummation of the large
logarithms which appear in the radiative corrections. We do not consider
 this question in this paper.
In future we shall continue to write only one argument $k_+$ as before
to avoid the complexity of the notation.

Using the time reversal invariance
of the strong interactions one can show that the functions $F^{\text{f,nf}}$
are real functions. As we shall see later this statement
is naturally realized in our models.

Consider the limit $N_c\rightarrow\infty$. As one can easily observe
\be{largeNc:Ci}
C_1\sim \mathcal{O}(N^0_c),\quad C_2\sim \mathcal{O}(N^{-1}_c).
\ee
But for the matrix elements:
\be{largeNc:me}
D^{\text{f}}\sim \tilde U^{\text{f}}\sim N_c,
\quad
D^{\text{nf}}\sim \tilde U^{\text{nf}}\sim N^{0}_c.
\ee
Hence both form factors $\alpha^{\text{f,nf}}$ are
of the same order with respect to large-$N_c$.
Note that in our analysis it is assumed that we first take the limit
 $m_Q\rightarrow\infty$ and after that $N_c\rightarrow\infty$.
 The conclusion is that despite the soft matrix elements have the different order
 with respect to large-$N_c$
  we must consider both contributions $\alpha^{\text{f}}$ and $\alpha^{\text{nf}}$.
 However the large-$N_c$ limit analysis is  useful because
 it allows to estimate the contributions to $\alpha^{\text{f}}$.

\subsection{  Form factor $\alpha^{\text{f}}$}
The corresponding matrix elements has the factorisable structure and
can be approximated  at the large-$N_c$ limit as the product of  two  matrix elements.
We have two non-perturbative functions corresponding to non-local
$d-$ \re{softme1} and $u-$quarks \re{softme2}. For the case of $d-$quark we can write
\be{1st}
\int\frac{d\lambda_{1}}{2\pi}~e^{i~k_{+}\lambda_{1}}
n^{\rho}\AS
\left\langle v^{\prime
}\left\vert \hb~\gamma_{\bot \sigma}P_{L}
~u~~~\bar{d}\left(  \lambda_{1}~n\right)
\gamma_\rho P_{L}~\Hv\right\vert v\right\rangle
\nonumber\\
 \simeq
\int\frac{d\lambda_{1}}{2\pi}~e^{i~k_{+}\lambda_{1}}
 \frac{1}{2}\langle v^{\prime}\vert
 \hb~\gamma_{\bot\sigma}P_{L}~u\vert 0~\rangle
 \quad
\langle 0\vert ~~\bar{d}\left(  \lambda_{1}~n\right)
  \gamma_+ P_{L}~\Hv \vert v\rangle
\\
 =\frac{1}{2}\left[  \frac{1}{2}(\varepsilon^{*\bot}_{D})_{\sigma}~F_{st}\right]
\left[  -\frac{1}{2}iF_{st}~\phi_{+}(k_{+})\right]  =-\frac{1}{8}iF_{st}%
^{2}~(\varepsilon^{*\bot}_{D})_{\sigma}~\phi_{+}(k_{+}),
\label{largeNc2}%
\ee
where  $F_{st}$ is the static mass-independent decay constant in HQET  which is related to the
physical constant of the heavy meson decay as
\be{Fst}
f_Q \sqrt{M_Q}=F_{st}(1+\mathcal{O}(\alpha_S)).
\ee
The function $\phi_{+}$ is known as $B-$meson light-cone distribution amplitude (LCDA )
\cite{NeGro}.
 Combining \re{largeNc2} with the parametrization \re{def:Df} one obtains
that at the large-$N_c$ limit
\be{Df:Nc}
\tilde{D}^{\text{f}}(k_{+})  & =&D^{\text{f}}(k_{+}),~\ ~\\
D^{\text{f}}(k_{+})  & =&\frac{1}{8}F_{st}^{2}~\phi_{+}(k_{+}).~
\ee
For the second  operator \re{softme2} one has
\be{2nd}
\int\frac{d\lambda_{2}}{2\pi}~e^{-il_+\lambda_{2}}
n^{\rho}{\AS}
\langle v^{\prime
},\varepsilon^*_{D}\vert \hb~\gamma_{\bot\sigma}%
P_{L}u\left(  \lambda_{2}~n\right)  ~~\bar{d}~\gamma_{\rho}P_{L}\Hv
\vert v \rangle
\nonumber \\
\overset{N_c\rightarrow\infty}{\simeq}\frac{1}{2}
\int\frac{d\lambda_{2}}{2\pi}~e^{-i l_+\lambda_{2}}
\langle v^{\prime},\varepsilon^*_{D}\vert \hb~\gamma^{\sigma
}P_{L}u\left(  \lambda_{2}~n\right) \vert 0~\rangle ~\langle
0\vert ~\bar{d}~\not n~P_{L}\Hv\vert v\rangle
 \\
 =
\frac{1}{2}\left[-\frac{i}{2}~F_{st}^{2}\right]
\frac{1}{2}\left( \varepsilon_{D\bot}^{*\sigma}~~g_{V}(l_{+})
-i\varepsilon_{\perp}^{\sigma\mu}
(\varepsilon^{*\bot}_{D})_\mu ~~g_{A}(l_{+})\right),
\label{UquarkNc}
\ee
where we introduced the transverse LCDAs:
\be{twist3}
\int\frac{d\lambda_{2}}{2\pi}~e^{-i~l_{+}\lambda_{2}}
\langle v^{\prime},
\varepsilon^*_{D}\vert \hb~\gamma_{\bot\sigma}P_{L}u\left(
\lambda_{2}~n\right)\vert 0
\rangle
=
\frac{1}{2}(\varepsilon^*_{D\bot})_{\sigma}~F_{st}~/v_{+}^{\prime}g_{V}(l_{+}/v_{+}^{\prime})
\nonumber\\
-\frac{i}{2}\varepsilon_{\perp\sigma\mu}(\varepsilon^*_{D})^{\mu}
~F_{st}/v_{+}^{\prime}~g_{A}(l_{+}/v_{+}^{\prime}).
\ee
 These  new
functions can be related to the LCDA $\phi_+$ due to the  heavy quark  spin-flavor
symmetry  \cite{largeM1, largeM2}.
 The corresponding relation reads, cf.\cite{NeGro}:
\be{gva:phi}
~g_{A}(l_{+})-g_{V}(l_{+})=-\phi_{+}(l_{+}). \ee Combining
\re{def:Uf},\re{UquarkNc} and \re{gva:phi} one finds
\be{UflNc}
\tilde{U}^{\text{f}}(l_{+})  & =\frac
{1}{8v_{+}^{\prime}}~F_{st}^{2}\phi_{+}(l_{+}/v_{+}^{\prime}).
\ee
Hence for the convolution integrals \re{conv:f} we obtain
\begin{equation}
~~\mathcal{D}^{\text{f}}=\int_{0}^{\infty
}dk_{+}~\frac{D^{\text{f}}\left(
k_{+}\right) ~}{~k_{+}}\simeq \frac{1}{8}F_{st}^{2}~\int_{0}^{\infty }dk_{+}~%
\frac{\phi _{+}(k_{+})~}{~k_{+}}=\frac{F_{st}^{2}}{8~\lambda _{B}}~,
\end{equation}%
\begin{equation}
\mathcal{U}^{\text{f}}=~\int_{0}^{\infty }dl_{+}\frac{\tilde{U}^{\text{f}%
}\left( l_{+}\right) }{l_{+}  }\simeq \frac{1}{%
8v_{+}^{\prime }}~F_{st}^{2}\int_{0}^{\infty }dl_{+}\frac{\phi _{+}(l_{+})}{%
l_{+}}=\frac{1}{v_{+}^{\prime }}\frac{F_{st}^{2}}{8~\lambda _{B}}=\frac{%
\mathcal{D}^{\text{f}}}{v_{+}^{\prime }}~.
\end{equation}
Substitution of these values into \re{af} gives
\be{af:Nc}
\alpha^{\text{f}}=\frac{C_2}{8}\frac{F_{st}^{2}}{\lambda _{B}}
\left \{ (\varepsilon_\gamma^*\cdot\varepsilon^*_D) \left(
e_d+\frac{e_u}{v_+^{\prime}} \right)
+i\varepsilon_{\perp\sigma\rho}\varepsilon_\gamma^{*\sigma}\varepsilon_D^{*\rho}
\left( e_d-\frac{e_u}{v_+^{\prime}} \right) \right \}.
\ee
The quantity $\lambda _{B}$ is well known from the phenomenology.
It was also estimated with the help of sum rules \cite{NeGro,BraunIK}.
For the
numerical estimate we accept the value  $\lambda _{B}(1\text{GeV})=0.35\pm
0.1$GeV. For the static decay constant we use the value \cite{Fst1,Fst2}
 $F_{st}(1\text{GeV})=0.35\pm 0.05$GeV$^{3/2}$ and for the coefficient function
 in the effective Hamiltonian  \re{Heff} we accept the leading order value
$C_2(m_b=4.8\text{GeV})=-0.268$ \cite{BurasC}. Then
\be{af:num} 10^{3}
\alpha^{\text{f}}\simeq
 (\varepsilon_\gamma^*\cdot\varepsilon^*_D)
\left(
0.97\text{GeV}^2
 \right)
+i\varepsilon_{\perp\sigma\rho}\varepsilon_\gamma^{*\sigma}\varepsilon_D^{*\rho}
\left( 7.1\text{GeV}^2 \right) .
\ee

\subsection{  Form factor $\alpha^{\text{nf}}$}
The two remaining non-perturbative functions $D^{\text{nf}}$ and $\tilde U^{\text{nf}}$
which contribute to the $\alpha^{\text{nf}}$ can be estimated using the method
of QCD sum rules.
For this purpose consider the following correlation functions (CFs)%
\be{corrf}
\bar{K}_{\bot q}^{\sigma\nu}\left(
\omega,\omega^{\prime},\lambda,v\cdot v^{\prime}\right)
&=&i\int dxdy~e^{-i(vx)\omega+i(v^{\prime}y)\omega^{\prime}%
}\left\langle 0\left\vert T\left\{  ~J^{\nu}_{D}(y),O_{q}^{\sigma}(\lambda
),J_{B}(x)\right\}  \right\vert 0\right\rangle,
\ee
where we used following notation%
\be{SRnot}
O_{d}^{\sigma}(\lambda)   =\frac{1}{2}\left[  \hb \gamma_{\bot}^{\sigma
}P_{L}\Hv \ \bar{d}(\lambda n)\gamma_+ P_{L}u-\hb\gamma_+ P_{L}\Hv \ \bar{d}(\lambda
n)\gamma_{\bot}^{\sigma}P_{L}u\right]  ,\\
O_{u}^{\sigma}(\lambda)   =\frac{1}{2}\left[  \hb \gamma_{\bot}^{\sigma
}P_{L}\Hv \ \bar{d}\gamma_+ P_{L}u(\lambda n)-\hb \gamma_+ P_{L}\Hv \ \bar{d}%
\gamma_{\bot}^{\sigma}P_{L}u(\lambda n)\right]  ,~\\
~~J_{D}(y)   =\bar{u}(y)\gamma^{\nu}h_{v'}(y),~\ \ \ \
J_{B}(x)=\bar{H}_v(x)i\gamma_{5}d(x).
\ee
The index $q=u,d$ is used to specify the non-local structure of the operator.
Each CF $\bar{K}_{\bot q}$ is parametrized by two form factors:
\be{def:Kud}
\bar{K}_q^{\sigma\nu}&=&\left(  g_{\bot}^{\sigma\nu}~-i\varepsilon_{\bot}^{
\sigma\nu} \right)  K_q+\left(  g_{\bot}^{\sigma\nu}~
+i\varepsilon_{\bot}^{  \sigma\nu }\right)  ~\tilde{K}_q~
\ee
Saturating the correlation functions with hadron states one obtains
for the relevant  form factors%
\be{Ku}
\tilde{K}_{u}  & =\frac{F_{st}^{2}~
}{4\left(  ~\bar{\Lambda}-\omega\right)  \left(  ~\bar{\Lambda}-\omega
^{\prime}\right)  ~} \frac{1}{2}\tilde{U} +~...~,
\\
K_{d}  & =\frac{F_{st}^{2} }{4\left(
~\bar{\Lambda}-\omega\right)  \left(  ~\bar{\Lambda}-\omega^{\prime}\right)
~} \frac{1}{2}D +~...~,
\label{Kd}
\ee
where dots denote the contributions from  higher resonances and continuum .

On the other hand for large negative $\omega,~\omega^{\prime}$ form factors
$K,\tilde{K}$ can be computed in Euclidian region.
\[
K_{d}=\int\frac{ds}{s-\omega}\int\frac{ds^{\prime}}{s^{\prime}-\omega^{\prime
}}\rho_{d}(s,s^{\prime},\lambda),~\ \ \ \ \tilde{K}_{u}=\int\frac{ds}%
{s-\omega}\int\frac{ds^{\prime}}{s^{\prime}-\omega^{\prime}}\tilde{\rho}%
_{u}(s,s^{\prime},\lambda),
\]
where spectral densities receive contributions from perturbation theory and
from vacuum condensates
\[
\rho=\rho^{\text{pert}}+\rho^{\text{cond}}%
\]
The leading order  diagrams for the perturbative and non-perturbative contributions are
shown in Fig.1.
\begin{figure}
[ptb]
\begin{center}
\includegraphics[
height=0.7558in,
width=5.3549in
]%
{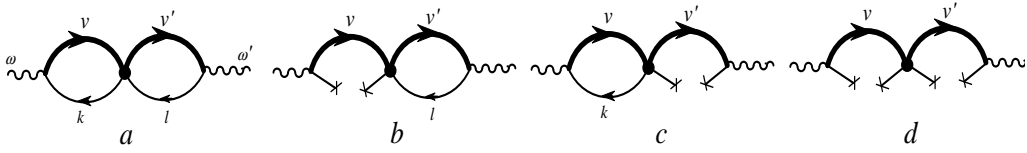}%
\caption{LO diagrams for the perturbative and non-perturbative spectral densities.
Diagrams with the gluon and quark-gluon operators are not shown. }%
\label{LOdiagram}%
\end{center}
\end{figure}
Performing the  subtraction of the the continuum contribution
($\omega_0$ is continuum threshold) and introducing the Borel
transformation with respect to $\omega$ and $\omega^{\prime}$ one obtains
\begin{align*}
\frac{1}{2}F_{st}^{2}~D(\lambda)~e^{-\bar{\Lambda}/t}  & =4\int_{0}%
^{\omega_{0}}ds\int_{0}^{\omega_{0}}ds^{\prime}~e^{-(s+s^{\prime})/2t}%
~\rho_{d}(s,s^{\prime},\lambda)~,\\
\frac{1}{2}F_{st}^{2}\tilde{U}(\lambda)~e^{-\bar{\Lambda}/t}  & =4\int
_{0}^{\omega_{0}}ds\int_{0}^{\omega_{0}}ds^{\prime}~e^{-(s+s^{\prime}%
)/2t}~\tilde{\rho}_{u}(s,s^{\prime},\lambda),~
\end{align*}
where we  accepted for
the values of the Borel parameters to be the same for the both channels
(the issue of the heavy quark symmetry):
\be{Borelt}
t_{1}=t_{2}=2t
\ee
and we also suppose that the value of the continuum threshold
$\omega_{0}$\ is the same as in two-point sum rules.
Performing Fourier transformation with respect to $\lambda$
 one obtains sum rules for the matrix elements in the
momentum space:%
\be{SRmom1}
\frac{1}{2}F_{st}^{2}D(k_{+})e^{-\bar{\Lambda}/t}
&\displaystyle  =4\int_{0}^{\omega_{0}%
}ds\int_{0}^{\omega_{0}}ds^{\prime}~e^{-(s+s^{\prime})/2t}~\rho_{d}%
(s,s^{\prime},k_{+})~,\\
\frac{1}{2}F_{st}^{2}\tilde{U}(l_{+})e^{-\bar{\Lambda}/t}
&\displaystyle  =4\int
_{0}^{\omega_{0}}ds\int_{0}^{\omega_{0}}ds^{\prime}~e^{-(s+s^{\prime}%
)/2t}~\tilde{\rho}_{u}(s,s^{\prime},l_{+}).~
\label{SRmom2}
\ee
The calculation of the diagrams in Fig.\ref{LOdiagram} provides
 the following analytical results for the spectral densities:
\be{rhou0}
\tilde{\rho}_{u}(s,s^{\prime},l_{+})
& =&-N_{c}\left(  \frac{1}{4\pi^{2}}\right)
^{2}\frac{1}{4v_{+}^{\prime}}~s^{2}\frac{l_{+}}{2v_{+}^{\prime}}~
\theta\left[ 0<l_{+}<2s^{\prime}v_{+}^{\prime}\right]
\nonumber \\
& +&\frac{\left\langle \bar{u}u\right\rangle }{16~\pi^{2}}\frac{1}%
{4v_{+}^{\prime}}\left[  ~\delta(s)\left[  1-\frac{1}{16}\frac{m_{0}^{2}%
}{4t^{2}}\right]  ~\frac{l_{+}}{2v_{+}^{\prime}}~\theta(0<l_{+}<2s^{\prime
}v_{+}^{\prime})  +s^{2}\delta(s^{\prime})~e^{-m_{0}^{2}~/64M^{2}%
}\delta\left(  \frac{l_{+}}{v_{+}^{\prime}}-\frac{m_{0}^{2}}{16t}\right)
\right]  ~
\nonumber \\
& -&\frac{N_{c}}{4v_{+}^{\prime}}\left(  \frac{\left\langle \bar{u}%
u\right\rangle }{4N_{c}}\right)
^{2}e^{-m_{0}^{2}~/64 t^{2}}~\delta(s)\left[
1-\frac{1}{16}\frac{m_{0}^{2}}{4t^{2}}\right]  ~~\delta(s^{\prime}%
)~~\delta\left(  \frac{l_{+}}{v_{+}^{\prime}}-\frac{m_{0}^{2}}{16t}\right)
\label{rhou}
\ee%
\be{rhod0}
\rho_{d}(s,s^{\prime},k_{+})&=&N_{c}\left(  \frac{1}{4\pi^{2}}\right)  ^{2}%
\frac{1}{4}~s^{\prime2}~\frac{k_{+}}{2}~\theta\left[  0<k_{+}<2s\right]
\nonumber \\
& +&\frac{N_{c}}{4}\left(  \frac{\left\langle \bar{u}u\right\rangle }{4N_{c}%
}\right)  ^{2}e^{-m_{0}^{2}~/64 t^{2}}~\delta(s^{\prime})\left[  1-\frac{1}%
{16}\frac{m_{0}^{2}}{4t^{2}}\right]  ~\delta(s)~\delta\left(  k_{+}%
-\frac{m_{0}^{2}}{16t}\right)
\label{rhod}
 \\
& -&\frac{\left\langle \bar{u}u\right\rangle }{16~\pi^{2}}\frac{1}{4}\left[
~\delta(s^{\prime})\left[  1-\frac{1}{16}\frac{m_{0}^{2}}{4t^{2}}\right]
\frac{~k_{+}}{2}\theta(0<k_{+}<2s)+~s^{\prime2}\delta(s)e^{-m_{0}^{2}
~/64 t^{2}}\delta\left(  k_{+}-\frac{m_{0}^{2}}{16t}\right)  \right]
\nonumber
\ee
The quantity $m_0$ is known as vacuum correlation length and defined as
$m_{0}^{2}=\left\langle \bar{q}g(\sigma G)q\right\rangle
/~\left\langle \bar{q}q\right\rangle \simeq0.8\text{GeV}^{2}$ The diagrams with quark
condensate in Fig.\ref{LOdiagram} have been computed using the technique of the
non-local condensate \ci{dubna1,dubna2}. In such approach one introduces
vacuum expectation value of the non-local operator
\begin{equation}
\left\langle 0\left\vert \bar{q}(x)[x,0]q(0)\right\vert 0\right\rangle
\simeq\left\langle 0\left\vert \bar{q}q\right\vert 0\right\rangle \int
_{0}^{\infty}d\nu~f(\nu)~e^{\nu x^{2}/4},
\end{equation}
which has to be understood as a model for the partial resummation
of the OPE to all orders. Such treatment allows to escape the
singular $\delta-$function terms which appear in the OPE with the
local condensates \cite{dubna2}. This is a very general situation
and it arises also in the sum rules  of the B-meson LCDA
\cite{NeGro, BraunIK}.
For the spectral function we accept the
simplest model suggested in \cite{dubna1,dubna2}:
\be{fnu}
f(\nu)=\delta(\nu-m_{0}^{2}/4).
\ee
Let us also remark that we neglect the terms with the gluon condensates because
corresponding contributions are small. The similar observation was made also in the sum
rules for B-meson LCDA \cite{NeGro, BraunIK}.

In the numerical calculations of sum rules
\re{SRmom1} and \re{SRmom2}
we  substitute the value of decay constant $F_{st}$ obtained from the corresponding the two-point sum
rule \cite{Fst1,Fst2}:
\be{SRFst}
\frac{1}{2}F^{2}(\mu)e^{-\bar{\Lambda}/t}=\frac{N_{c}}{2\pi^{2}}\int
_{0}^{\omega_{0}}ds~s^{2}e^{-s/t}-\frac{1}{2}\left\langle \bar{u}%
u\right\rangle \left[  1-\frac{m_{0}^{2}}{16t^{2}}\right]  ,
\ee

It is instructive to consider the so-called local duality
limit $t\rightarrow\infty$. Then the sum rules expressions  are
simplified and one obtains
\be{LD:sr1}
~D^{\text{nf}}(k_{+})\bigr |_{t\rightarrow\infty}&=&\frac{\omega_{0}^{2}}{8\pi^{2}}
\frac{k_{+}}{2\omega_{0}}~
\left(  1-\frac{k_{+}}{2\omega_{0}}\right)
\theta\left[  ~0<k_{+}<2\omega_{0}\right],
\\
\tilde{U}^{\text{nf}}(l_{+})\bigr |_{t\rightarrow\infty}  & =&  -
\frac1{v_{+}^{\prime}}
\frac{\omega
_{0}^{2}}{8\pi^{2}}\frac{l_{+}}{2\omega_{0}v_{+}^{\prime}%
}~\left(  1-\frac{l_{+}}{2\omega_{0}v_{+}^{\prime}}\right)
\theta\left[~0<l_{+}<2\omega_{0}v_{+}^{\prime}\right]
\ee
As one can see, the both functions are localized in the region $k_+<2\omega_0 v_+$.
 We expect that this is  valid only for the  leading order approximation,
 similar to the $B-$meson LCDA \ci{BraunIK}. Another important property is the ``good''
behavior in the limit  $k_+\rightarrow 0$. Such behavior at small values of the
momentum fraction do not contradict
to the existence of the convolution integrals \re{conv:nf}.

In the numerical estimates we use for the Borel mass $t$ and continuum
threshold $\omega_{0}$ the same values as in the the
two-point sum rules \cite{Fst1}
\be{prmts}
0.3~\text{GeV} <t<0.6~\text{GeV,}
\quad
\omega_{0} =0.8-1.0~\text{GeV},
\ee
and $\left\langle \bar{u}u\right\rangle =-(240\text{MeV})^{3}$,$~m_{0}^{2}%
\simeq0.8\text{GeV}^{2}$.
From the expressions for the spectral densities \re{rhou} and \re{rhod} one can
easily  find that
\be{SR:UDnf}
\mathcal{D}^{\text{nf}}  =\int\frac{dk_{+}}{k_{+}}D^{\text{nf}}(k_{+}),\,\,
\mathcal{U}^{\text{nf}}  =\int\frac{dl_{+}}{l_{+}}\tilde{U}^{\text{nf}%
}(k_{+})=-\mathcal{D}^{\text{nf}}/v_{+}^{\prime},
\ee
with
$v_{+}^{\prime}=M_B/M_D=2.63$.
Therefore we  provide the
results only for one quantity $\mathcal{D}^{\text{nf}} $.
Numerically we obtained
\be{SR:Dnf} \mathcal{D}^{\text{nf}}
=\left(  1.32\pm0.16 \right)  \times10^{-2}~\text{GeV}^{2},
\ee
where the uncertainty arises from the $t-$ and
$\omega_0-$variations. Hence assuming \re{SR:UDnf} we obtain
\be{afn:Nc} \alpha^{\text{nf}}\simeq {C_1}\mathcal{D}^{\text{nf}}
\left\{ (\varepsilon_\gamma^*\cdot\varepsilon^*_D) \left(
e_d-\frac{e_u}{v_+^{\prime}} \right)
+i\varepsilon_{\perp\sigma\rho}\varepsilon_\gamma^{*\sigma}\varepsilon_D^{*\rho}
\left(
 e_d+\frac{e_u}{v_+^{\prime}}
\right) \right \}.
\ee
Substituting the leading order value for the coefficient function
$C_1(m_b=4.8\text{GeV})=1.12$ one has
\be{anf:num}
10^{3} \alpha^{\text{nf}}\simeq
 (\varepsilon_\gamma^*\cdot\varepsilon^*_D)
\left(
-8.64\text{GeV}^2
 \right)
+i\varepsilon_{\perp\sigma\rho}\varepsilon_\gamma^{*\sigma}\varepsilon_D^{*\rho}
\left( -1.20\text{GeV}^2 \right) .
\ee
Comparing this result with
the analogous expression for $\alpha^{\text{f}}$ \re{af:num} we observe that both
form factors $\alpha^{\text{f}}$ and $\alpha^{\text{nf}}$,
from the factorizable and non-factorizable soft matrix
elements are of the same order. From the structure of expressions \re{af:num} and \re{anf:num}
it is easy to see that factorizable contribution
dominates in the physical form factor $F_1\simeq 3.8 \times 10^{-9}\text{GeV}$ but the non-factorizable
term provides the largest contribution to the
the second  physical  form factor $F_2\simeq -2.5 \times 10^{-9}\text{GeV}$.

\subsection{Branching fraction estimate and conclusions}
With the above results we can estimate the branching fraction.
Using for the CKM matrix elements $|V_{ud}|=0.974$ and
$|V_{cb}|=0.415$ and $\alpha=1/137$ we obtain for the branching ratio
\be{bran}
\mathcal{B}(\bar{B}^{0}\rightarrow D^{\ast0}\gamma)=(1.52\pm 0.35)\times10^{-7}%
\ee
The uncertainty given in \re{bran} originate from the
uncertainties of the hadronic matrix elements. We observe that our
estimate is of two order magnitude smaller that the
experimental bound
$\mathcal{B}(\bar{B}^{0}\rightarrow D^{\ast0}\gamma)<2.5\times 10^{-5}$ \cite{BABAR}.
Our estimate is also significantly smaller than the values provided by previous
considerations \cite{theor1,theor2,theor3}.  The main conclusion is
that such small quantity most probably can not be measured at
existing $B$-factories.

On the other hand our estimate has to be considered carefully. We
have used only the leading order contribution. There are a lot of
corrections which a priory may be of considerable size. We did
not consider the resummation of the possible large Sudakov
logarithms associated with the choice of the factorization scale.
Typically, the large corrections  arise also from the
next-to-leading contributions to the jet functions in the SCET
approach because the corresponding hard scale  $\sim \bar \Lambda
E_\gamma$ is not very large \cite{NLOSCETII}. The complication
also arises due to the fact that there are two different  heavy
quark masses $m_{b}$ and $m_{c}$ that introduce an additional
scale ambiguity. Therefore on the background of these remarks our
result has to be considered only as a leading order qualitative
estimate. However, we expect that all effects mentioned above can
not provide
 such strong enhancement that can make the value of the branching measurable
for BABAR or BELLE experiments.

\subsection*{Acknowledgments}
The author is grateful to K.~Semenov-Tian-Shansky for careful reading of the manuscript.
The
work is supported
by the Sofja Kovalevskaja Programme of the Alexander von Humboldt
Foundation, the Federal Ministry of Education and Research and the
Programme for Investment in the Future of German Government.


\end{document}